\documentclass[showpacs,preprintnumbers,showkeys,preprint,prb,floatfix]{revtex4}

\usepackage{epsfig}
\usepackage{amsmath}
\usepackage{amssymb}
\usepackage{bm}

\begin{document}

\title{Crack Growth Laws from Symmetry}

\author{James P. Sethna}
\affiliation{Laboratory of Atomic and Solid State Physics, Clark Hall, Cornell
University, Ithaca, NY 14853-2501}
\email[Corresponding author, Electronic address:]{sethna@lassp.cornell.edu}
\homepage{http://www.lassp.cornell.edu/sethna/sethna.html}
\date{\today}

\begin{abstract}
Reviewing work done in collaboration with Jennifer Hodgdon, we derive
the most general crack growth law allowed by symmetry for mixed-mode
three-dimensional fracture. We do so using the system developed in
condensed matter physics to derive new laws that emerge on long length
and time scales. In our derivation, we provide a symmetry interpretation
for the three modes of fracture, we rederive the law giving the crack
growth direction in two dimensional fracture, and we derive a growth law
appropriate for three dimensional simulations. We briefly discuss
related work subsequent to ours, incorporating disorder, dynamics, and
other internal variables.
\end{abstract}


\maketitle

\def\ddx#1#2{ {\frac{\partial #2}{\partial #1}} }
\def\kI{K_{\rm I}}
\def\kII{K_{\rm II}}
\def\kIII{K_{\rm III}}
\def\nhat{\hat n}
\def\bhat{\hat b}
\def\that{\hat t}
\def\Xvec{\vec X}
\def\Rb{R_b}
\def\Rt{R_t}
\def\dadotb#1#2{ \frac{\partial #1}{ \partial s} \cdot #2 }

In this paper, I review work with Jennifer Hodgdon\cite{Hodgdon} using
symmetry principles to derive the growth law for mixed-mode, curved cracks
for isotropic materials in three dimensions. In the process, I'll attempt
to illustrate the general approach that condensed-matter physicists have 
developed to derive new phenomenological laws for macroscale physical phenomena.

The physical laws governing the behavior of many physical systems can
be derived using symmetry. Basically, there are five steps, which are
not necessarily applied sequentially:

\noindent{\it I. Pick an order parameter field.} The order parameter
field is a state variable, which at a point $\bf x$ summarizes the
current state of the material in the local neighborhood of $\bf x$.

\noindent{\it II. Use the symmetries of the problem.} 

\noindent{\it III. Imagine writing the most general possible law.} 

\noindent{\it IV. Simplify the theory using small parameters, power
counting, etc.} Usually one will need to expand in gradients of the
order parameter: the theory then describes only behavior at long
wavelengths. In dynamical evolution laws, one will naturally also expand
in time derivatives, specializing to low frequencies. One often expands
in powers of the order parameter.

\noindent{\it V. Solve the theory.} In many cases, this is straightforward,
analytically or computationally. In other cases temperature, dirt, or noise
introduce fluctuations that remain important on all length scales (such
as near continuous phase transitions), and renormalization-group methods
may be needed.

Landau introduced this system to derive the forms of free energies and
near-equilibrium dynamics of materials with broken symmetries. His
methods have been applied extensively to exotic liquid crystals,
superconductors and superfluids, magnetic systems, ferroelectrics,
incommensurate systems, martensites, and in spatially extended dynamical
systems. In many of these systems, the phenomenological Landau theory
was effectively used long before a microscopic mechanism or model was
developed. More recently, essentially the same procedure has been used to
derive evolution equations in systems which are far from equilibrium.

How shall we proceed to apply these ideas to cracks?

\bigskip\bigskip
\noindent {\bf I. Pick an order parameter field.}
\medskip

\begin{figure}[thb]
\epsfig{file=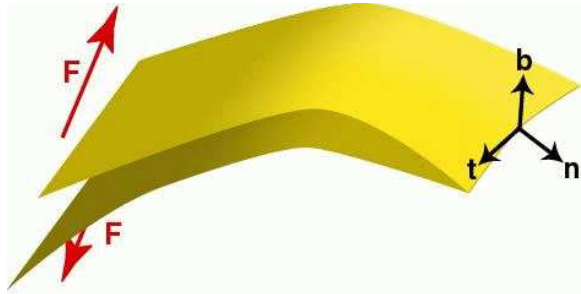, width=3truein}
\caption{\label{CrackTemplate} 
{\bf Order parameter for crack growth.}
The crack is parameterized by arclength $s$; the crack edge is a curve 
$\Xvec(s)$ growing in direction $\nhat(s)$. The tangent to the crack
edge $\that=d\Xvec/ds$ and the normal to the crack plane
$\bhat=\that\times\nhat$ are derivable from $\Xvec$ and $\nhat$.}
\end{figure}

\noindent
The region of interest in fracture is the immediate vicinity of the crack front:
in general, a curved line in space (figure \ref{CrackTemplate}). Let us
parameterize this curve by arc
length $s$, so the curve is $\Xvec(s)$. The geometry of the crack edge
demands not only the coordinates of this line, but also the orientation of
the crack surface as it approaches the line. We can define a unit tangent
vector $\that = d\Xvec/ds$ to the crack edge, the direction of crack
growth $\nhat$ (perpendicular to $\that$), and the normal to the crack plane
$\bhat=\that \times \nhat$. The functions $\Xvec(s)$ and $\nhat(s)$ are enough
to determine the geometry of the crack; we use them as the order parameter
for the problem.

Of course, the crack growth rates will depend strongly on the applied 
stresses at the crack front. These depend on the loading far away, as well
as the shape of the crack surface behind the growing crack front. In special
circumstances (nearly straight crack fronts in 
mode~I\cite{Cotterell,RamanathanFisher1}) we can write closed-form 
expressions for these stresses
in terms of integrals over $\Xvec(s)$ and $\nhat(s)$; more generally these
stresses can be calculated from elastic theory, {\it e.g.} using finite
element analysis \cite{IngraffeaWeb,Wawrzynek}. We thus assume that
the stresses as a function of $s$ are given.

\bigskip\bigskip
\noindent {\bf II. Use the symmetries of the problem.}
\medskip

\noindent
We assume that our material is isotropic, and will consider only
quasi-static fracture. On length scales short compared to the curvatures
of the crack and compared to the gradients of the
stresses at the crack front, we have two independent discrete symmetries
for the local crack geometry: reflection $\Rb$ in the plane of the crack
(taking $\bhat$ to $-\bhat$) and reflection $\Rt$ in the plane
perpendicular to the crack front $\that$. There is a third symmetry,
a 180$^\circ$ rotation about the axis $\nhat$, which is the product of
the two other symmetries.

We can use these symmetries first to simplify the characterization of
the quasistatic stresses at the crack tip.  As for all linear
problems with symmetries, we may decompose a general solution of the
linear elastic problem into solutions whose displacement (and strain)
fields are odd or even under the two symmetries $R_b$ and $R_t$. 
For an uncracked material whose strain is constant along $\that$, 
a general elastic solution can be decomposed into multipoles: for 
each power $n$ of the distance $r$ to the $\that$ axis, there are six 
elastic solutions whose strains vary as $r^n$. (For example, there
are six solutions with constant strain $n=0$, corresponding to the six
independent coefficients of the elastic strain tensor.)
There are three solutions which are even in $R_b$ and $R_t$, and one each
of the other three possibilities. For the medium with a crack, three of these
elastic solutions are not allowed by the condition that there be no
traction at the crack surface. Instead, there are three new classes of
solutions with strain fields depending on half-integer powers of $r$,
with non-zero displacements across the crack 
surface\cite{Ruina,HodgdonThesis}.

What are all these elastic solutions? The solutions which involve
strains varying as $r^n$ for $n>0$ are large at the boundaries of the
sample but vanish quickly at the crack front. They are important for
solving for elastic deformations in complex geometries, but are
irrelevant for crack growth except when the sample size is comparable to
the size of the nonlinear zone at the crack front. The solutions which 
have $n\le -1$ have diverging total energy at the crack tip. These solutions
are important for matching the boundaries of the nonlinear zone to the
elastic theory, but will die away at distances comparable to the size
of the nonlinear zone. Their magnitudes thus are determined by the 
local geometry and stress at the crack front, and hence can also be ignored.
We are left with three ordinary elastic strains $n=0$ and three 
strain fields with crack opening displacements $n=-1/2$. These latter fields are
the familiar three modes of fracture (table \ref{tab:FractureModes}).

\begin{table}
\begin{tabular}{|c|c|c|c|}
\hline
&$\Rb$ & $\Rt$ & \\
\hline
\vbox to 1.2truein{\vfil\hbox{Mode I, $K_{I}$}\vfil}
&\vbox to 1.2truein{\vfil\hbox{Even}\vfil}  
&\vbox to 1.2truein{\vfil\hbox{Even}\vfil} 
& \epsfig{file=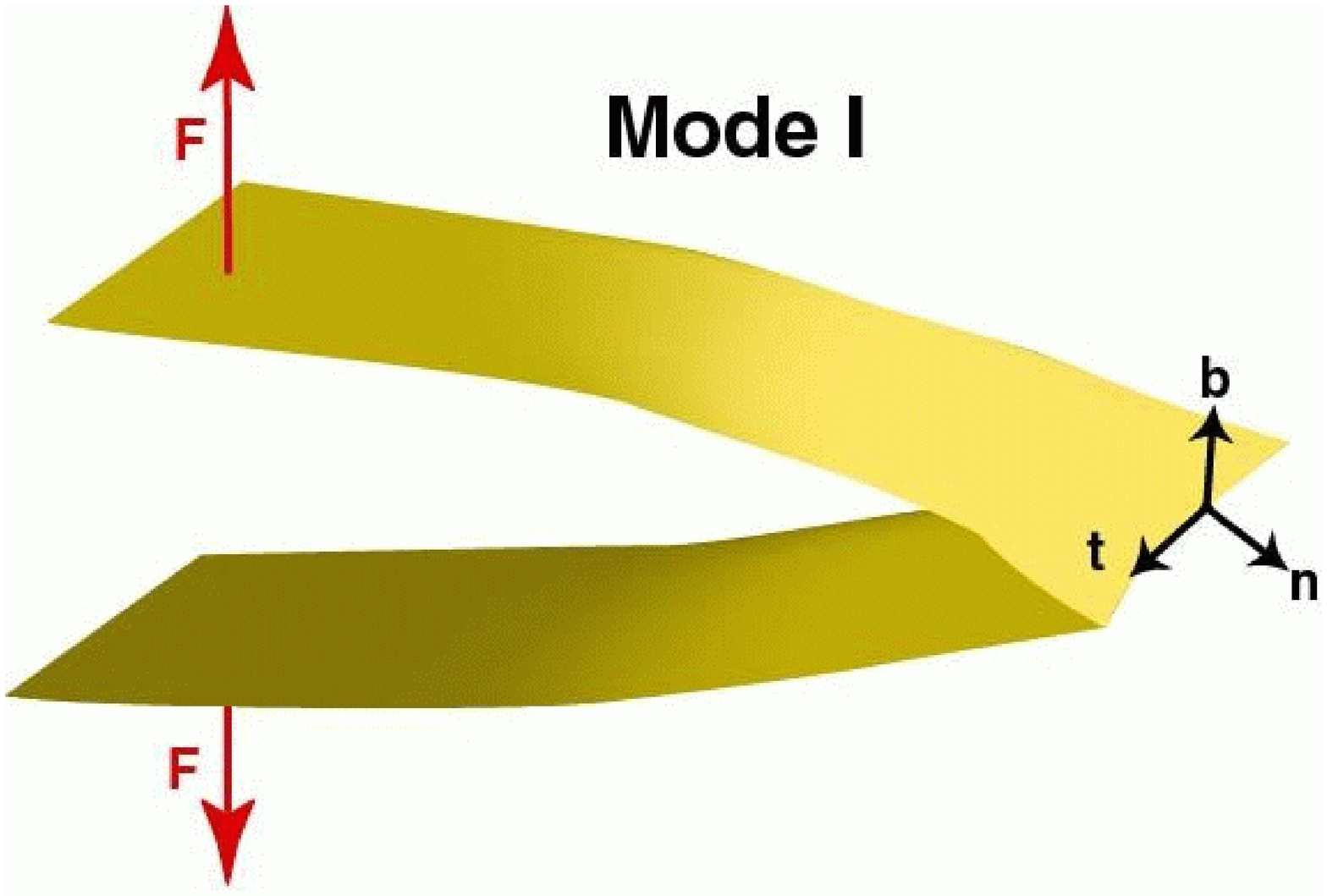, width=1.5truein}\\
\hline
\vbox to 0.6truein{\vfil\hbox{Mode II, $K_{II}$}\vfil}
&\vbox to 0.6truein{\vfil\hbox{Odd}\vfil}  
&\vbox to 0.6truein{\vfil\hbox{Even}\vfil} 
& \epsfig{file=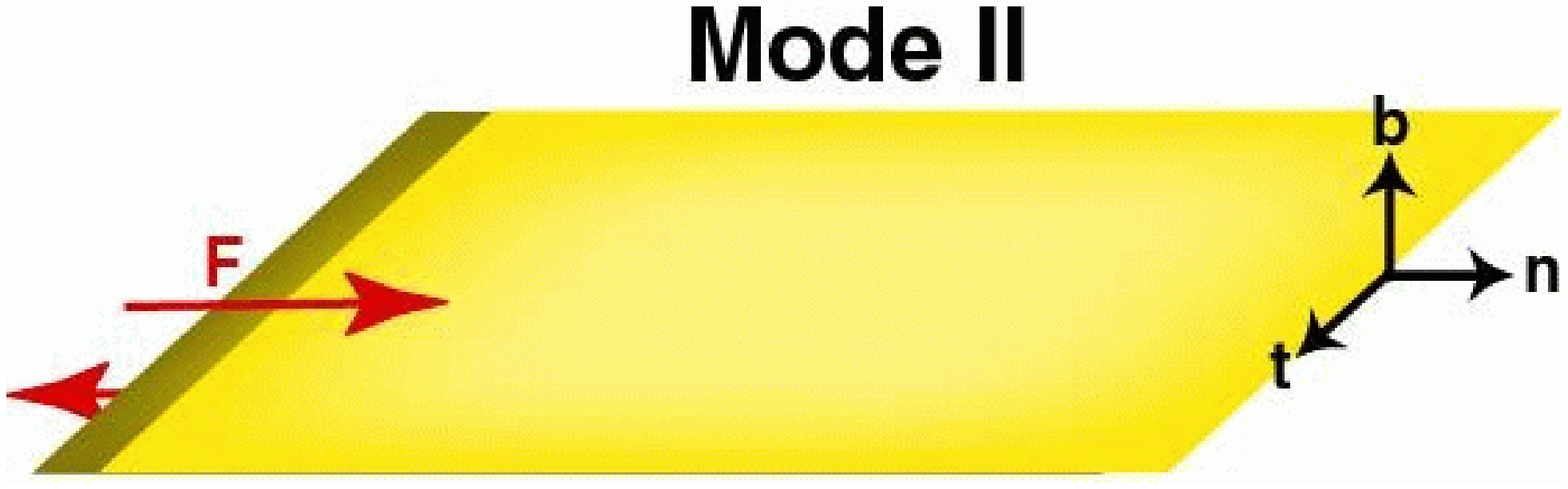, width=1.5truein}\\
\hline
\vbox to 0.7truein{\vfil\hbox{Mode III, $K_{III}$}\vfil}
&\vbox to 0.7truein{\vfil\hbox{Odd}\vfil}  
&\vbox to 0.7truein{\vfil\hbox{Odd}\vfil} 
& \epsfig{file=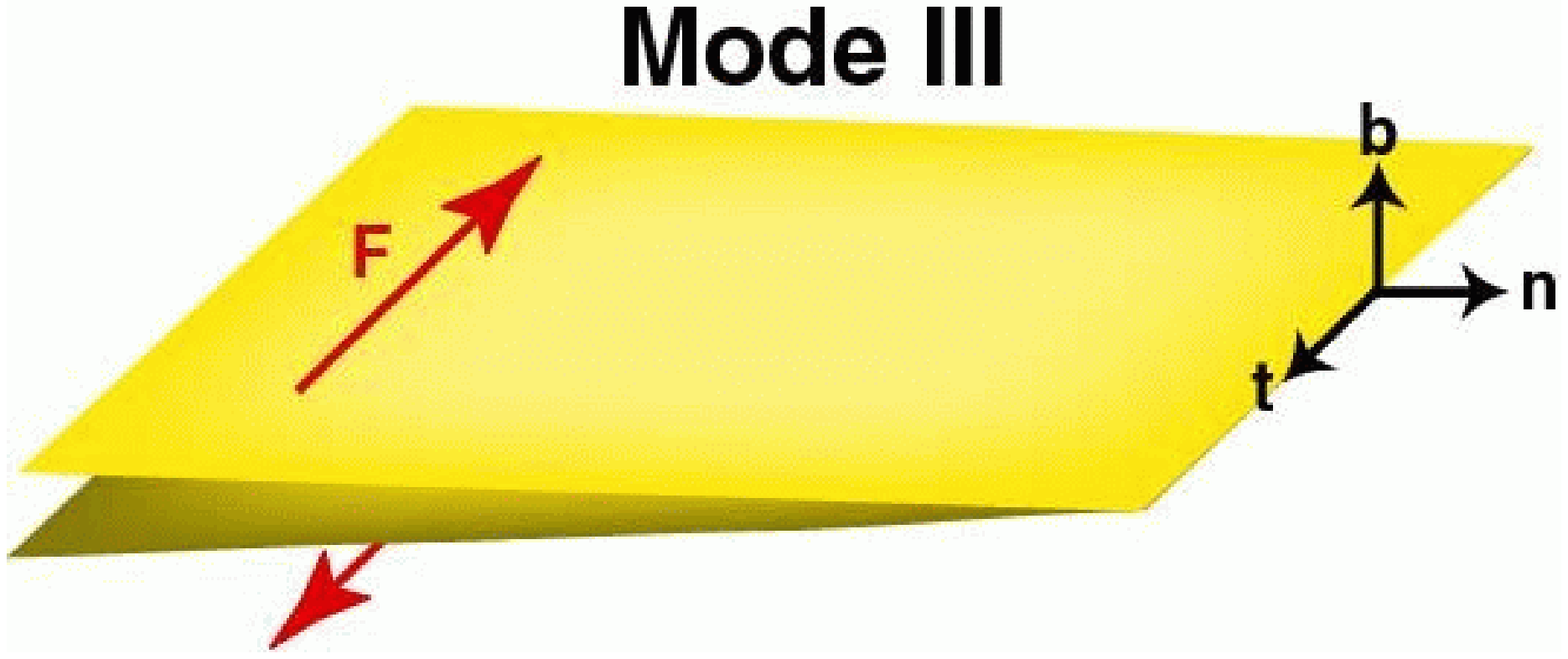, width=1.5truein}\\
\hline
&Even & Odd & No crack opening\\
\hline
\end{tabular}
\caption{\label{tab:FractureModes}Symmetries and Modes of Fracture.}
\end{table}

In our earlier work\cite{Hodgdon}, we ignored the $n=0$ elastic 
stresses near the crack tip. It's likely that their effects are relatively
small, since they aren't intensified near the crack tip, and we will ignore
them in this review as well. It would be straightforward to incorporate
the overall strains into the equations of motion, for a more complete theory.

There is one more symmetry in our problem: an artificial symmetry which
is introduced by our theoretical description. It is not always
convenient to use arc-length $s$ to parameterize the crack. In
particular, it makes the growth laws nonlocal: even for a stationary
portion of the crack, if a far-away region of the crack with smaller $s$
changes in length the functions $\Xvec(s)$ and $\nhat(s)$ will shift
sideways. Of course, we can use any parameterization that's convenient.
Changing parameterizations in a formal sense is analogous to changing
the choice of gauge in electromagnetism \cite{SteveLanger}. For roughly
straight cracks parallel to an axis $z$, we would likely use $z$-gauge
with $\Xvec(z)$ and $\nhat(z)$; for loops we would likely use
$\theta$-gauge. Changing the equation of motion from one gauge to
another is called a {\it gauge transformation}. There is thus a {\it
gauge symmetry}: the physical evolution of the crack must be unchanged,
or {\it gauge invariant}, when we change from one parameterization to another.

\begin{figure}[thb]
\epsfig{file=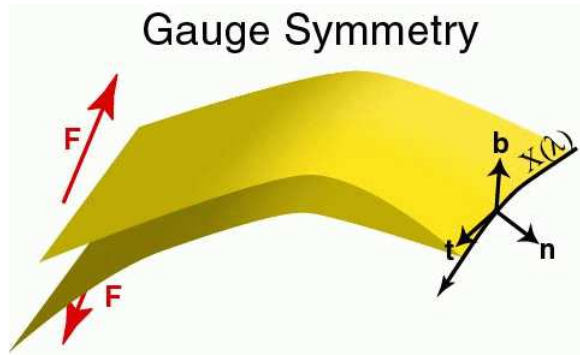, width=3truein}
\caption{\label{CrackGauge} 
One must choose a parameterization of the crack edge, $\lambda$ (choosing
a gauge). The form of the equations of motion will change for different 
choices for the parameterization, but physical quantities like the location
of the crack front must be unchanged (gauge invariant).}
\end{figure}

\bigskip\bigskip
\noindent {\bf III. Write the most general possible law.}
\medskip

\noindent
In two-dimensional fracture, we can immediately write down the most
general possible form for the crack velocity and turning rate. The
position $\Xvec$ of the crack moves forward along $\nhat$ with velocity
$v$ which depends on the stress intensity factors. The velocity does
not change sign under $\Rb$ or $\Rt$, so $v$ cannot have terms odd in
$K_{II}$ or $K_{III}$, giving us the first law
\begin{equation}
\partial {\bf X}/\partial t = v(K_I,K_{II}^2,K_{III}^2) {\bf \hat n}.
\label{eq:2Dv}
\end{equation}
Similarly, the rate of rotation of the crack surface changes sign under
reflection in the plane of the crack $\Rb$, and doesn't change sign under
$\Rt$ so it must have one factor of $K_{II}$ times an arbitrary function
of $K_{I}$, $K_{II}^2$, and $K_{III}^2$:
\begin{equation}
\partial {\bf \hat n}/\partial t = -f(K_I,K_{II}^2,K_{III}^2)
                        K_{II} {\bf \hat b}.
\label{eq:2Df}
\end{equation}

In three dimensions, we must incorporate possible dependences of the 
equations of motion on the local curvatures of the crack front and the
gradients of the stress intensity factors. To linear order in gradients,
Hodgdon found
\begin{eqnarray}
 \partial x / \partial t =& v \nhat + w \that \cr
 \partial \nhat / \partial t =& - \left[ \ddx{s}{v} +
   w \ddx{s}{\that}  \cdot \nhat \right] \that +
  \left[ - f \kII + g_{\rm I} \kIII \ddx{s}{\kI} +
   g_{\rm II}\kII \kIII \ddx{s}{\kII} + g_{\rm III}\ddx{s}{\kIII} + \right. \cr
&\qquad \left.  h_{tb} \dadotb{\that}{\bhat} +
   h_{nt} \kII \dadotb{\nhat}{\that} +
 (h_{nb}\kII\kIII + w)  \dadotb{\nhat}{\bhat} \right] \bhat ,\cr
\label{eq:3D}
\end{eqnarray}
where $f$, $g_{\alpha}$, and $h_{ij}$ are functions of $\kI$, $\kII^2$, and
$\kIII^2$, and the velocity $v$ can be a function of these and a number of
gradient terms\cite{Hodgdon}.

\bigskip\bigskip
\noindent {\bf IV. Simplify the theory using small parameters, 
power counting, etc.}
\medskip

\noindent
By focusing on quasi-static fracture, and by assuming that the
equations of motion involve only first derivatives in time, we're already making
use of the small ratio of the velocity of the crack growth to the
natural material velocities (velocity of sound, surface relaxation, etc.)
By confining our attention to the external strains that go as $r^{-1/2}$
and $r^0$ we made the assumption
that the nonlinear zone of the crack is small compared to the system size
(so-called K-dominant fracture, since the three stress intensity factors
dominate). In three dimensions, by keeping terms to linear order in gradients,
we have again assumed the nonlinear zone is small compared to curvatures
and changes in stresses.

We can use the small size of the nonlinear zone a third time to establish the
relative sizes of different terms in our equations of motion. In our
two dimensional equations of motion (\ref{eq:2Dv}) and (\ref{eq:2Df}),
the functions $v$ and $f$ have different units: $v/K f$ has units of
length, where $K$ is any one of the stress intensities. It is natural to
assume that this length will be set by the size of the nonlinear zone,
and in any case it is part of our approach to assume that this length is small.

\bigskip\bigskip
\noindent {\bf V. Solve the theory.} 
\medskip

\noindent
It is well known that cracks loaded with a mixture of mode~I and more~II
will turn abruptly. In our formulation, the function $f$ governs the
rate at which the crack turns. Hodgdon used results of Cotterell and
Rice \cite{Cotterell} to show that if the angle of the crack differs
from the angle that makes $K_{II}=0$ by a small amount $\Delta \theta$,
then $K_{II} = K_I \Delta \theta/2$. Combined with the two growth laws
above, we find that $\Delta \theta \sim \exp(-f K_I x/2 v)$, the crack
turns to pure mode~I exponentially with a material-dependent decay
length of $2 v / f K_I$: precisely twice the characteristic length scale
we assumed was small! Thus we derive the {\it principle
of local symmetry}\cite{GS}, which says that a mode~II fracture will
turn abruptly (that is, on a length comparable to the nonlinear zone
size) until it is pure mode~I.

There were two other rules in the literature for picking the crack growth
direction: one maximizing the energy release\cite{Wu}, and one moved in
the direction of minimum strain energy density\cite{Sih}. Cracks grown
by these different rules (move forward by a small step size $\Delta x$, 
recalculate the stress intensity factors, repeat) all gave rather similar
predictions for the shapes. Our analysis did not make any assumptions about
microscopic mechanisms, so we can apply it to a hypothetical material which
behaved according to these other crack growth rules.
We conclude that these other rules will end up forcing the crack to turn
until $K_{II}=0$, yielding in all cases the principle of local symmetry!
Instead of a nonlinear zone size, the turning radius is given by the 
step size of the algorithm $\Delta x$. Our analysis predicts that {\it all
three growth rules are equivalent} in the limit of small stepsize/turning
radius.

\bigskip
\noindent {\bf Where to go from here?}
\medskip

\noindent
Hodgdon's analysis is now a decade old. What came next?

First, Hodgdon used the three-dimensional growth law (equation \ref{eq:3D})
to investigate the stability of crack growth under mixtures of mode~III
and mode~I. All of the new 3D materials functions $g_X$, $h_{yz}$, and $w$ in
equation (\ref{eq:3D}) multiply gradients, so they aren't unusually large 
like $f$.
Hodgdon found that mixed-mode fracture can be stable or unstable depending
on details: in particular, for $g_I>0$ steady-state mode~III cracks are
unstable to small perturbations. Mode~III fracture is unstable experimentally
to the formation of a ``factory roof'' morphology, with ramps followed by
cliffs. With the computers at that time, we couldn't get into the nonlinear
regime needed to test whether our theory leads to the same jagged solutions
as seen experimentally: Hodgdon's linear stability analysis remains
unpublished\cite{HodgdonThesis}.

Second, real materials are dirty: not only impurities and inclusions,
but also the random fracture strengths introduced by polycrystallinity
suggest the incorporation of randomness in the evolution equations.
Introducing randomness is also motivated by the experimental observation
of Bouchaud\cite{Bouchaud}, who finds roughness on all scales, with
power-law height-height correlation functions. Ramanathan and
Fisher\cite{RamanathanFisher1} (see also \cite{Nattermann}) studied the
effects of incorporating disorder into a dynamics similar to that
described here. They discovered that it does indeed predict a fracture
surface which is rough on all length scales, but with much weaker
randomness than that observed experimentally (logarithmic rather than
power law). It seems likely that the experimental roughness, at least in
ductile and intergranular fracture, reflects void growth and the
presence of grains and inclusions\cite{Anderson}.

Thirdly, many effects of dynamics on crack growth have been ignored in
our discussion. The theory as described here is appropriate, perhaps, for 
fatigue crack growth where mass and inertia are not important.  New
phenomena arise as the crack speeds up (such as the mirror-mist-hackle 
morphology transitions\cite{Fineberg}). Incorporating the inertial
dynamics of the growing front does lead to interesting traveling wave
solutions\cite{FrontWaves}. 

Finally, there is the likely possiblity that there are other important
degrees of freedom that are important, and need to be incorporated into the
order parameter describing the current state of the crack tip. 
There are recent indications that the curvature of the crack tip\cite{Argonne}
and the effects of surface tension and diffusion\cite{LobkovskyKarma,Brener}
may be important.

\begin{acknowledgments}

This work was supported by NSF KDI-9873214 and NSF ACI-0085969\/.
We thank Nick Bailey for helpful discussions.

\end{acknowledgments}


\begin{thebibliography}{99}

\bibitem{Hodgdon} Jennifer A. Hodgdon and James P. Sethna, {\sl Phys. Rev. B}
{\bf 47}, 4831 (1993), James P. Sethna, invited talk at the 10th International
Conference on Fracture, December 2001.

\bibitem{Cotterell} B. Cotterell and J. R. Rice, {\sl Int. J. Frac.}
{\bf 16}, 155 (1980).

\bibitem{RamanathanFisher1} S. Ramanathan, D. Ertas, and D. S. Fisher,
{\sl Phys. Rev. Lett.} {\bf 79}, 873 (1997).

\bibitem{IngraffeaWeb} Cornell Fracture Group, http://www.cfg.cornell.edu/.

\bibitem{Wawrzynek} P. A. Wawrzynek and A. R. Ingraffea, {\it Discrete
Modeling of Crack Propagation: Theoretical Aspects and Implementation Issues
in Two and Three Dimensions}, Ph.D. Thesis, Cornell University, 1991.

\bibitem{Ruina} C. Y. Hui and Andy Ruina, {\it Int. J. Frac.} {\bf 72}
97 (1995).

\bibitem{HodgdonThesis} J. A. Hodgdon, {\it Three-Dimensional Fracture:
Symmetry and Stability}, Ph. D. Thesis, Cornell University, 1993.

\bibitem{SteveLanger} S. A. Langer, R. E. Goldstein, and D. P. Jackson,
{\sl Phys. Rev. A} {\bf 46}, 4894 (1992).

\bibitem{GS} R. V. Gol'dstein and R. L. Salganik {\sl Int.
J. Frac.} {\bf 10}, 507 (1974).

\bibitem{Wu} C. H. Wu, {\sl J. Appl. Mech.} {\bf 45}, 553 (1978).

\bibitem{Sih} G. C. Sih, p. xv in {\it Mechanics of Fracture 2: Three 
Dimensional Crack Problems} (eds. M. K. Kassir and G. C. Sih), Noordhoof
International, Netherlands, 1975.

\bibitem{Nattermann} P. F. Arndt and T. Nattermann, ``A New Criterion for 
Crack Formation in Disordered Materials'', cond-mat/0012113, accepted
for publication in {\sl Phys. Rev. B.}

\bibitem{Bouchaud} E. Bouchaud, {\sl J. Phys. Condens. Matter} {\bf 9},
4319 (1993) and references therein.

\bibitem{Anderson} T. L. Anderson, {\it Fracture Mechanics, Fundamentals
and Applications}, CRC Press, Boca Raton, 1991, chapter 5.

\bibitem{Fineberg} J. Fineberg, S. P. Gross, M. Marder, and H. L. Swinney,
{\sl Phys. Rev. Lett.} {\bf 67}, 457 (1991).

\bibitem{FrontWaves} S. Ramanathan and D. S. Fisher, {\sl Phys. Rev. Lett.}
{\bf 79}, 877 (1997); J. R. Rice, Opening Lecture at the 20th International
Congress of Theoretical and Applied Mechanics, Chicago, Aug. 27, 2000.

\bibitem{Argonne} I. S. Aranson, V. A. Kalatsky, V. M. Vinokur, {\sl Phys.
Rev. Lett.} {\bf 85}, 118 (2000).

\bibitem{LobkovskyKarma} A. Lobkovsky and A. Karma, ``Tip splitting
instability in a phase field model of mode III dynamic fracture'', APS
meeting presentation Q27.002, March 2002; A. Lobkovsky, condensed-matter
seminar, Cornell University, spring 2002.

\bibitem{Brener} Efim A. Brener, Robert Spatschek, ``Fast crack propagation
by surface diffusion'', cond-mat/0204056.

\end{thebibliography}
\end{document}